\documentclass[amsmath,amssymb, aps, preprint,superscriptaddress]{revtex4-2}

\usepackage[utf8]{inputenc}
\DeclareUnicodeCharacter{2212}{-}
\usepackage{times}%
\usepackage{graphicx}
\usepackage{amsmath}
\usepackage{color}
\usepackage{natbib}
\usepackage{footnote}
\usepackage{bm}
\usepackage[version-1-compatibility]{siunitx}
\usepackage{upgreek}
\usepackage[colorinlistoftodos]{todonotes}
\usepackage{hyperref}
\hypersetup{colorlinks=true, citecolor=blue, linkcolor=blue}
\usepackage{soul}
\usepackage{comment}
\usepackage{setspace}

\makeatletter
\newcommand*{\rom}[1]{\expandafter\@slowromancap\romannumeral #1@}
\makeatother

\begin{document}
\title{Low-density superconductivity in SrTiO$_3$ bounded by the adiabatic criterion}
\author{Hyeok Yoon}
\thanks{Present address: Department of Physics, University of Maryland, College Park, MD 20742}
\email{hyoon13@umd.edu}
\affiliation{Stanford Institute for Materials and Energy Sciences, SLAC National Accelerator Laboratory, Menlo Park, California 94025, USA}
\affiliation{Geballe Laboratory for Advanced Materials,Stanford University, Stanford, California 94305, USA}
\author{Adrian G. Swartz}
\affiliation{Stanford Institute for Materials and Energy Sciences, SLAC National Accelerator Laboratory, Menlo Park, California 94025, USA}
\affiliation{Geballe Laboratory for Advanced Materials,Stanford University, Stanford, California 94305, USA}
\author{Shannon P. Harvey}
\affiliation{Stanford Institute for Materials and Energy Sciences, SLAC National Accelerator Laboratory, Menlo Park, California 94025, USA}
\affiliation{Geballe Laboratory for Advanced Materials,Stanford University, Stanford, California 94305, USA}
\author{Hisashi Inoue}
\affiliation{Stanford Institute for Materials and Energy Sciences, SLAC National Accelerator Laboratory, Menlo Park, California 94025, USA}
\affiliation{Geballe Laboratory for Advanced Materials,Stanford University, Stanford, California 94305, USA}
\author{Yasuyuki Hikita}
\affiliation{Stanford Institute for Materials and Energy Sciences, SLAC National Accelerator Laboratory, Menlo Park, California 94025, USA}
\affiliation{Geballe Laboratory for Advanced Materials,Stanford University, Stanford, California 94305, USA}
\author{Yue Yu}
\affiliation{Stanford Institute for Materials and Energy Sciences, SLAC National Accelerator Laboratory, Menlo Park, California 94025, USA}
\affiliation{Geballe Laboratory for Advanced Materials,Stanford University, Stanford, California 94305, USA}
\author{Suk Bum Chung}
\affiliation{Department of Physics and Natural Science Research Institute, University of Seoul, Seoul 02504, Republic of Korea}
\author{Srinivas Raghu}
\email{sraghu@stanford.edu}
\affiliation{Stanford Institute for Materials and Energy Sciences, SLAC National Accelerator Laboratory, Menlo Park, California 94025, USA}
\affiliation{Geballe Laboratory for Advanced Materials,Stanford University, Stanford, California 94305, USA}
\author{Harold Y. Hwang}
\email{hyhwang@stanford.edu}
\affiliation{Stanford Institute for Materials and Energy Sciences, SLAC National Accelerator Laboratory, Menlo Park, California 94025, USA}
\affiliation{Geballe Laboratory for Advanced Materials,Stanford University, Stanford, California 94305, USA}

\date{\today}

\begin{abstract}
 SrTiO$_3$ exhibits superconductivity for carrier densities $10^{19}-10^{21}$ cm$^{−3}$.Across this range, the Fermi level traverses a number of vibrational modes in the system, making it ideal for studying dilute superconductivity. We use high-resolution planar-tunneling spectroscopy to probe chemically-doped SrTiO$_3$ across the superconducting dome. The over-doped superconducting boundary aligns, with surprising precision, to the Fermi energy crossing the Debye energy. Superconductivity emerges with decreasing density, maintaining throughout the Bardeen-Cooper-Schrieffer (BCS) gap to transition-temperature ratio, despite being in the anti-adiabatic regime. At lowest superconducting densities, the lone remaining adiabatic phonon van Hove singularity is the soft transverse-optic mode, associated with the ferroelectric instability. We suggest a scenario for pairing mediated by this mode in the presence of spin-orbit coupling, which naturally accounts for the superconducting dome and BCS ratio.
\end{abstract}

\maketitle

\newpage

The occurrence of superconductivity in electron-doped SrTiO$_3$ has been a mystery ever since its discovery over a half-century ago \cite{Schooley:1964}. To date, it remains one of the lowest density known bulk superconductors, only recently surpassed by Bi \cite{Prakash:2017}. It was also the first material shown to exhibit a doping-dependent superconducting dome \cite{Koonce:1967}. This has been observed in many subsequent materials, for which the dome structure is often a result of nontrivial electron interactions. By contrast, electrons in SrTiO$_3$ are expected to reflect more simple, semiconductor-like charge degrees of freedom. The lattice is unusual for a semiconductor, in that SrTiO$_3$ exhibits a soft transverse-optic phonon mode that nearly condenses at low temperature \cite{Yamada:1969}. The absence of an ordered ferroelectric ground state is attributed to quantum fluctuations \cite{Muller:1979}.

There has been a renaissance of interest in this problem from a number of fronts. The ability to engineer low-dimensional superconductivity in various SrTiO$_3$ heterostructures has emerged \cite{Caviglia:2008,Kozuka:2009,Cheng:2015}, where the low carrier density is highly amenable to electrostatic control. Furthermore, there is increasing experimental evidence for a close relationship between ferroelectricity and superconductivity \cite{Rischau:2017,Ahadi:2019,Tomioka:2019}, developed in the broader context of modern studies of bulk single crystals and thin films. Theoretically there has been progress in treating electron-phonon interactions in the regime of strong coupling \cite{Chubukov:2020, Kumar:2021}, scenarios for plasmon and plasmon-polariton coupling \cite{Takada:1980,Ruhman:2016}, and considerations of quantum critical electrical polarization fluctuations \cite{Appel:1969,Edge:2015}. We take advantage of many of these developments in our work here.

\begin{figure}[ht]
\centering
\includegraphics[width=160 mm]{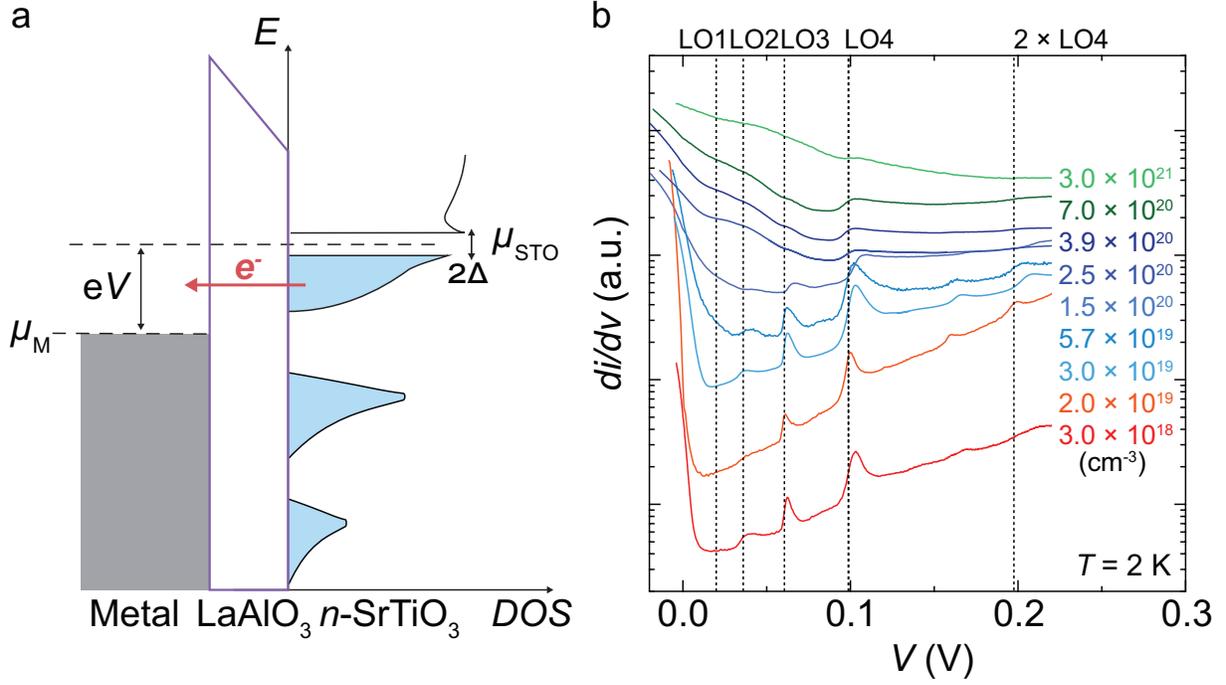}
\caption{Probing the electronic structure of electron-doped SrTiO$_3$ using planar tunnel junctions. (a) Schematic junction, with the density of states (DOS) indicated on the right for the superconducting state (with gap $\Delta$). LaAlO$_3$ is used as an epitaxial polar tunnel barrier between the SrTiO$_3$ and the metallic counter-electrode, for which the chemical potentials are $\mu_\textrm{STO}$ and $\mu_\textrm{M}$, respectively. When a positive bias-voltage $V$ is applied to the electrode, electrons in occupied SrTiO$_3$ states tunnel across the barrier to unoccupied states in the electrode. (b)	Normal-state differential tunneling conductance $di/dv$ measured at $T = 2$ K for varying carrier density $n$ (derived from normal state Hall measurements). The vertical dotted lines indicate e$V$ equal to longitudinal optic phonon energies LO1, LO2, LO3, LO4, and 2 $\times$ LO4.
\label{fig1}}
\end{figure}

We study the systematic evolution of the electronic structure of doped SrTiO$_3$ using planar tunnel junctions as shown in Fig. \ref{fig1}(a). We utilize high-temperature synthesis of highly-crystalline, substitutionally-doped thin-film SrTiO$_3$ in the step-flow growth regime and atomic-scale polar tunnel barriers (two or three unit cell LaAlO$_3$) covered by a high work-function metal \cite{Swartz:2018}. These structures have been developed to suppress carrier depletion at the SrTiO$_3$ surface, enabling flat-band measurements of the electronic structure down to low carrier densities \cite{Yajima:2015}. Representative differential conductance ($di/dv$) measurements performed at 2 K are shown in Fig. \ref{fig1}(b) for Nb doping from 0.02 - 2 atomic \%, and 3.5 and 7 atomic \% La doping above the bulk solubility limit of Nb. As can be seen in the forward bias spectra, clear longitudinal-optic (LO) phonon replicas are observed, most prominently for the highest-energy LO4 mode and at the lowest densities. These features can be analyzed to estimate \cite{Swartz:2018} the electron-phonon coupling $\lambda_\textrm{{LO4}}$, ranging from $\lambda_\textrm{LO4} > 1$ for the lowest densities, and rapidly diminishing with increasing density and screening. The phonon peak linewidth is also a good probe of the Fermi energy $E_\textrm{F}$ at the interface \cite{Swartz:2018}, which confirms the expected doping evolution for flat-band conditions.

\begin{figure*}[ht]
    \centering
\includegraphics[width=160 mm]{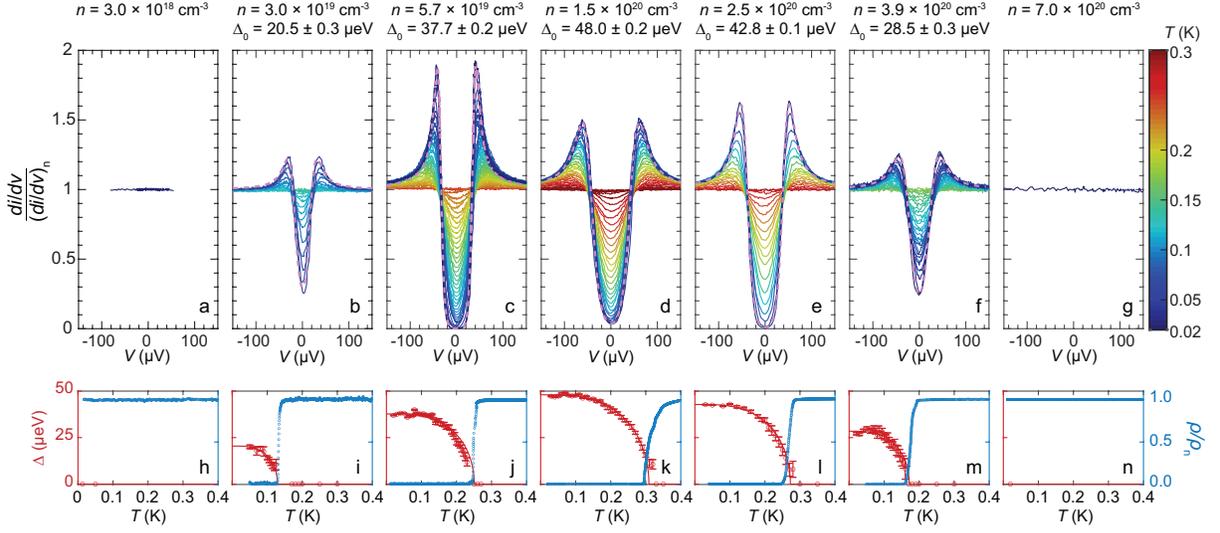}
\caption{Tunneling gap spectra for various carrier densities $n$ in SrTiO$_3$.
(a)-(g): Tunneling conductance $di/dv$ normalized to the normal-state tunneling conductance ($di/dv)_\textrm{n}$, for various temperatures (following the color scale on the right) and densities (increasing from left to right) across the superconducting state. Fits with the BCS gap function are given by the magenta dashed lines for the base temperature spectra. (h)-(n): Temperature dependent gap $\Delta$ and  resistivity $\rho$ normalized to the normal-state resistivity ($\rho_\textrm{n}$) are shown for the left (red) and right (blue) axis, respectively. Error bars for $\Delta$ denote the 95 \% confidence interval.
\label{fig2}}
\end{figure*}

Figure \ref{fig2} shows $di/dv$ measured across the doping series at various temperatures in a dilution refrigerator, normalized to its normal state value $(di/dv)_\textrm{n}$. The samples span the superconducting dome, as well as non-superconducting ground states at high and low doping.

   The lower panels of Fig. \ref{fig2} show the resistive transition and the temperature dependent superconducting gap $\Delta(T)$. We extract $\Delta$ at each temperature by fitting the Bardeen-Cooper-Schrieffer (BCS) gap function via 
   
   \begin{equation}
   \begin{aligned}
    \frac{\left. di/dv \right|_V}{(di/dv)_\textrm{n}|_V} &= - \int \nu(E) \left. \frac{\partial f_\textrm{FD}}{\partial E} \right|_{E+eV} dE, \\
    \nu(E) &= \mathbf{Re} \left[\frac{E-i\Gamma}{\sqrt{(E-i\Gamma)^2-\Delta^2}}\right],
    \end{aligned}
    \label{eq1}
    \end{equation}
    where $f_\textrm{FD}$ is the Fermi-Dirac function, $\nu$ is the density of states, and $\Gamma$ is the Dynes quasiparticle broadening parameter.  From this we can find the zero temperature gap $\Delta_0$ from the extrapolated fitting of $\Delta (T)$, 
    
   \begin{equation}
   \Delta (T) = \Delta_0 \textrm{tanh} \left[\frac{\pi}{1.76}\sqrt{\frac{2}{3}1.43\left(\frac{T_\textrm{c}}{T}-1\right)}\right],
   \label{eq2}
   \end{equation}
   where $\Delta_0$ and the superconducting transition temperature $T_\textrm{c}$ are free parameters of the fit \cite{Devereaux:1995}. We find that $T_\textrm{c}$ deduced from gap spectroscopy is very close to the value determined by 50$\%$ of the resistive transition, as shown in Fig. \ref{fig2}(h)-(n). In particular, no evidence of a pseudogap is observed, in contrast to one- and two-dimensional superconductivity in SrTiO$_3$ heterostructures, which may reflect their enhanced quantum fluctuations, classical phase fluctuations, and sensitivity to disorder \cite{Cheng:2015, Richter:2013, Chen:2018} 
   
\begin{figure}[ht]
\centering
\includegraphics[width=160 mm]{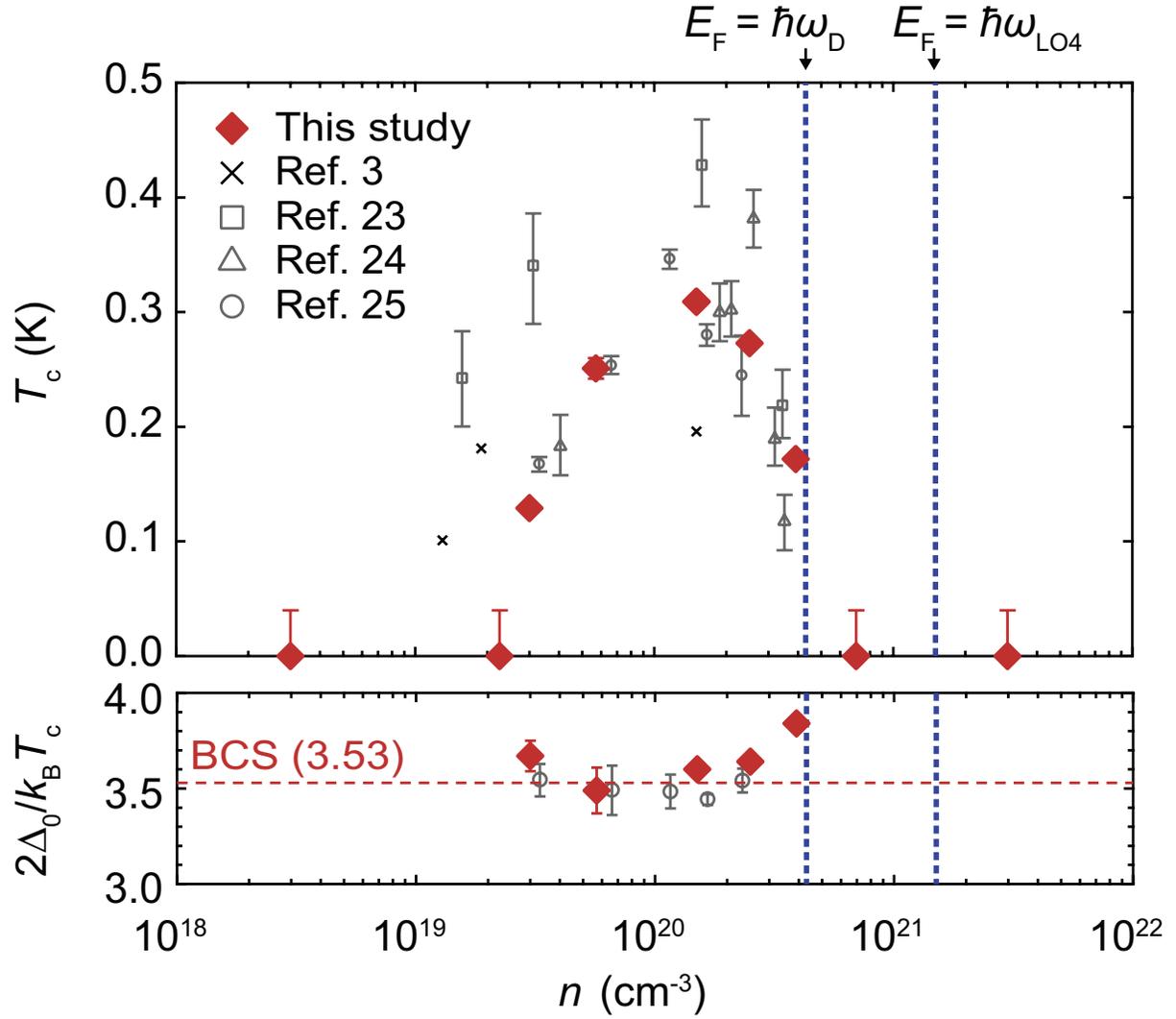}

\caption{$T_\textrm{c}$  and  $2 \Delta_0 / k_\textrm{B} T_\textrm{c} $ across the SrTiO$_3$ superconducting dome. Data extracted from Fig. 2 (red diamonds). The vertical dotted lines indicates $n$ for which $E_{\text{F}}=\hbar \omega_\textrm{D}$, and $E_{\text{F}}=\hbar \omega_\textrm{LO4}$,
where $E_{\text{F}}(n)$ and $\hbar \omega_\textrm{D}$ are derived from \cite{McCalla:2019}. The gray markers show the resistive transition from other experiments \cite{Koonce:1967,Lin:2014,Collignon:2017,Thiemann:2018}. The horizontal dashed red line indicates the BCS value $2 \Delta_0 / k_\textrm{B} T_\textrm{c}= 3.53$.
\label{fig3}}
\end{figure}

   Figure \ref{fig3}(a) plots $T_\textrm{c}$ as a function of doping across the superconducting dome, which is consistent with prior studies of the resistive transition \cite{Koonce:1967, Lin:2014,Collignon:2017,Thiemann:2018}. Also indicated in Fig. \ref{fig3}(a) is the carrier density $n$ at which $E_\textrm{F}$ is equal to the Debye energy $\hbar\omega_\textrm{D}$ extracted from specific heat measurements \cite{McCalla:2019}. It is quite striking that this density corresponds to the overdoped boundary for superconductivity in SrTiO$_3$. The LO4 (zone center) phonon energy $\hbar \omega_\textrm{LO4}$ is crossed by $E_\textrm{F}$ at even higher carrier density (Fig. \ref{fig3}(a)).
   Conventional superconductivity in metals is mediated by the retarded attractive interaction arising from electron-phonon coupling. Both the original weak-coupling BCS limit and the extension to strong-coupling in Migdal-Eliashberg theory \cite{Migdal:1958,Eliashberg:1960} are based on the adiabatic condition that $\hbar \omega_\textrm{D}/E_{\text{F}} \ll 1$. Here we find that superconductivity in SrTiO$_3$ only emerges in the anti-adiabatic regime. Despite this, we find that $2 \Delta_0/ k_\textrm{B} T_\textrm{c} $ (where $k_\textrm{B}$ is the Boltzmann constant) conspicuously mantains a value near the weak-coupling BCS limit of 3.53 across the entire superconducting dome, as shown in Fig. \ref{fig3}(b), consistent with recent microwave conductivity measurements \cite{Thiemann:2018}.

\begin{figure}[ht]
\centering
\includegraphics[width=160 mm]{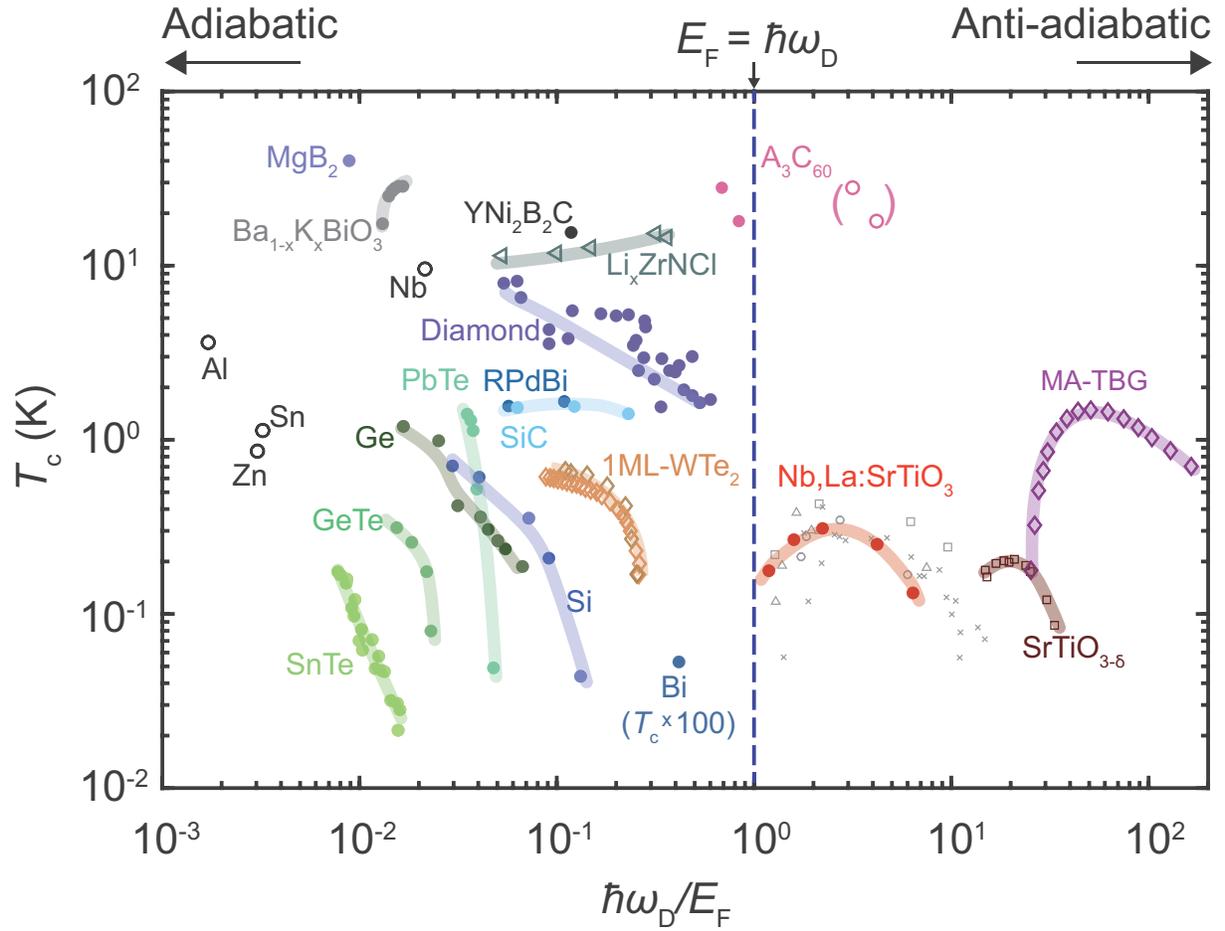}
\caption{$T_\textrm{c}$ versus $\hbar\omega_\textrm{D}$/$E_\textrm{F}$ for various superconductors.
Dots, triangles, and diamonds denote 3D, quasi-2D, and 2D materials, respectively. Guidelines for each material are drawn if the tuning parameter is electron/hole doping. A$_3$C$_{60}$ for intramolecular phonons and 
both normal-state (filled circles) and superfluid (open circles with parentheses) densities are shown. RPdBi: R=Y, Lu, and for A$_3$C$_{60}$: A=K, Rb, from right to left. The vertical dashed line denotes the adiabatic criterion  $\hbar\omega_\textrm{D}$/$E_\textrm{F} = 1$. Materials information and references are given in Supplemental Table S1. 
\label{fig5}}
\end{figure}

  To place this surprising observation in context, we show in Fig. \ref{fig5} the variation of $T_\textrm{c}$ as a function of $\hbar \omega_\textrm{D}/E_\textrm{F}$ estimated for a wide range of superconductors, with emphasis on those that occur at low carrier density and for which phonon contributions to pairing are established or possible. This includes semiconductors with the face-centered-cubic diamond structure (C, Si, SiC, Ge) and the rock-salt structure (PbTe, SnTe, GeTe). In all of these cases, superconductivity is observed only for the adiabatic condition.  Despite the extremely low carrier density ($\sim$ 10$^{17}$ cm$^{-3}$) in superconducting Bi, it is also adiabatic due to its very low effective mass. Superconductivity in the fullerides \cite{Gunnarsson:1997} is difficult to categorize given the very different energy scales for intra/inter-molecular phonons. Indeed the possible relevance of anti-adiabatic pairing motivated theoretical investigations of this problem \cite{Pietronero:1995,Grimaldi:1995, Chakravarty:1992}. 
  
  The only superconductors clearly found in the anti-adiabatic regime are doped SrTiO$_3$, magic-angle twisted bilayer graphene (MA-TBG), and heavy fermion systems (not shown). For completeness we also plot in Fig. \ref{fig5} the second superconducting dome observed in SrTiO$_{3-\delta}$ exclusively for electron-doping via oxygen vacancies. There are open questions whether this second dome occurs due to inhomogeneities arising from vacancy clustering \cite{Szot:2002, Bretz-Sullivan:2019}; experimentally we cannot controllably access this regime in our tunneling structures either via direct synthesis of SrTiO$_{3-\delta}$ \cite{Ohtomo:2007} or reduction (given the LaAlO$_3$ barrier).

\begin{figure}[ht]
\centering
\includegraphics[width=140 mm]{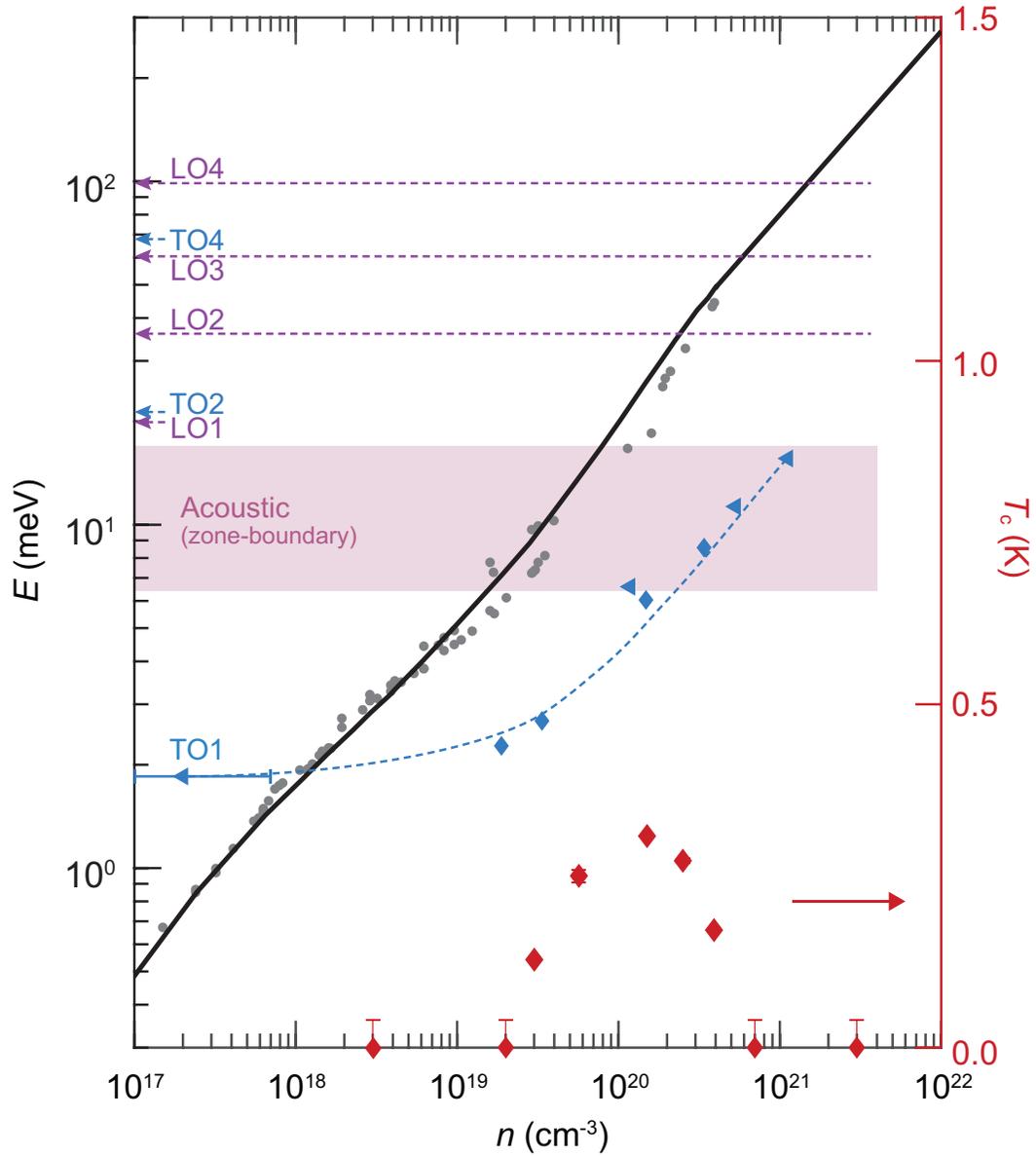} 
\caption{Variation of $E_\textrm{F}$, the TO1 phonon, and the superconducting dome with carrier density in SrTiO$_3$. Gray circles show $E_\textrm{F}$ measured via quantum oscillations \cite{Lin:2015}, and the black solid line shows a scaled electronic structure fit to various measurements \cite{vanderMarel:2011}. The energies of zone-center optical phonons are indicated (extended as dashed lines for those that are constant with doping following Fig. \ref{fig1}(b)), as well as the energy range of zone-boundary acoustic phonons (shaded region) \cite{Choudhury:2008}. The variation of the TO1 phonon energy with doping is plotted with blue triangles \cite{Bauerle:1980} and  diamonds \cite{Mechelen:2008}.
\label{fig4}}
\end{figure}

 Going beyond the average measure of the phonon energy given by $\hbar \omega_\textrm{D}$, additional insight can be gained by considering the various phonon modes crossed by $E_\textrm{F}$ with decreasing density, as shown in Fig. \ref{fig4}. In the low-density half of the superconducting dome, $E_\textrm{F}$ is lower than all but one of the optical phonon modes of the system. Although  $E_\textrm{F}$ crosses the zone-boundary acoustic modes in this region, we note that they cannot mediate pairing across the small electron pockets at the zone center in SrTiO$_3$. Therefore, other than the extremely low phonon density of states arising from the linear acoustic branches, the only prominent phonon density remaining is the anomalously soft transverse optic phonon mode TO1.
 
The experiments reported here have several consequences for proposed electron-phonon pairing mechanisms in SrTiO$_{3}$ which,  broadly construed,  fall into two categories.  
	The first set of ideas involve anti-adiabatic pairing scenarios, where the entire spectrum of phonons, both of longitudinal and transverse modes, is involved in providing an attractive interaction.  Since the Debye energy is large compared to the Fermi energy, the resulting pairing problem is in an inverted regime where the electrons are slow compared to the phonons.  In this framework, one would not {\it{a priori}} expect a BCS value for $2 \Delta_0/ k_\textrm{B} T_\textrm{c} $, which crucially depends on the adiabatic limit to apply.  Furthermore, we expect substantial mass renormalization to occur in this regime due to the breakdown of Migdal-Eliashberg theory \cite{Chubukov:2020}.  The absence of such effects casts doubt on the relevance of anti-adiabatic pairing in light of the experimental measurements. 

	The second set of ideas involves a conventional BCS pairing mechanism, often in the context of ferroelectric quantum criticality, where the phonon scale is small compared to $E_\textrm{F}$ \cite{Kedem:2018,Kozii:2019,Gastiasoro:2020}.  Given the low carrier concentration, the only viable candidate for such pairing is using the TO1 mode.  Without spin-orbit coupling, the electrons can only couple to pairs of TO1 phonons, as noted in Ref. \cite{vanderMarel:2019}.  However, with spin-orbit coupling, pairing can be mediated by single TO1-phonon processes (via scattering between distinct orbitals), and can naturally give rise to a BCS value for the gap to $T_\textrm{c} $ ratio. 

	A further advantage of the TO1-mediated pairing scenario is that it can also account for a superconducting dome occurring at low Nb concentrations. To see why, note first that with niobium doping $x$, the effective spring constant acquires a density dependent contribution of the form
	\begin{equation}
	    K \approx K_0 + K_1 x, \ \ \  x \ll 1
	\end{equation}
   where $K_0$ is the TO1 mode  spring constant associated with Ti \cite{Bauerle:1980}, and $K_1$  can be derived in the spirit of an effective medium approximation where the system is viewed as a binary alloy.
   It suggests that the hardening of the TO1 mode from Nb doping stems from substitution rather than from an electronic mechanism.  Furthermore, with the adiabatic assumption, where the TO1 mode entirely generates the pairing interaction, the effective BCS pairing eigenvalue is given by
    \begin{equation}
    \lambda_\textrm{BCS} \propto \frac{x^{1/3}}{K_0 + K_1 x}.
    \end{equation}
     At low concentrations, the pairing strength decreases due to the vanishing of the density of states, whereas at larger concentrations, the weakening of the effective interaction due to the hardening of the TO1 phonon causes the attenuation of the pairing strength. Finally, invoking the Coulomb pseudopotential, $\mu^*$, which is certainly valid within the adiabatic regime, we find that the transition temperature $T_c \sim \exp{\left[-1/(\lambda - \mu^* ) \right]}$ decays with density in a convex fashion and abruptly vanishes when $\lambda = \mu^*$. Our hypothesis here is further bolstered by the fact that the LO modes are hardly affected with Nb doping (Fig. \ref{fig1}), which would suggest that their role in superconducting properties is relatively limited. 

Thus within this simple framework, we are able to rationalize the observation of a superconducting dome, along with the observed BCS $2 \Delta_0/ k_\textrm{B} T_\textrm{c} $ ratio. It is interesting to consider how the dome in bulk SrTiO$_3$ relates to that found in electrostatically-gated LaAlO$_3$/SrTiO$_3$  \cite{Caviglia:2008, Chen:2018} given the differences in the density of states, orbital structure, and the presence of Rashba coupling at the interface. Finally we note that the results and discussion presented here are likely relevant to the recent discovery of $T_\textrm{c} \sim 2$ K at KTaO$_3$ interfaces \cite{Liu:2021, Chen:2021}, which has a much higher spin-orbit coupling scale than for SrTiO$_3$, but somewhat reduced proximity to ferroelectricity.

   
\clearpage
\section*{Figures.}

\clearpage 
\section*{Acknowledgements}
We thank K. Behnia, S. A. Kivelson, I. Fisher, and Y. Suzuki for useful discussions. 
This work was supported by the Department of Energy, Office of Basic Energy Sciences, Division of Materials Sciences and Engineering, under contract No. DE-AC02-76SF00515; and by the Gordon and Betty Moore Foundation’s Emergent Phenomena in Quantum Systems Initiative through Grant GBMF9072 (dilution refrigerator equipment and A.G.S.). S.B.C. was supported by the National Research Foundation of Korea (NRF) grants funded by 
the Korea government (MSIT) (2020R1A2C1007554) and the Ministry of Education 
(2018R1A6A1A06024977).

%

\end{document}


\begin{center}
\fontsize{18}{12}\selectfont Supplementary Materials for \break 

\fontsize{14}{12}\selectfont Low-density superconductivity in SrTiO$_3$ bounded by the adiabatic criterion. \break 

\fontsize{12}{12}\selectfont Hyeok Yoon, Adrian G. Swartz, Shannon P. Harvey, Hisashi Inoue, Yasuyuki Hikita, Yue Yu, Suk Bum Chung, Srinivas Raghu, Harold Y. Hwang \break 

{\fontsize{12}{12}\selectfont Correspondence to: hyoon13@umd.edu, sraghu@stanford.edu, hyhwang@stanford.edu}

\end{center}
This PDF file includes: \newline
Table S1  \newline
Caption for Table S1 

\clearpage

\textbf{Table. S1}

  \begin{table}[h!]
  \small
  \centering
  \begin{tabular}{| c | c | c | c |}
    \hline
    Materials & Carrier type &  $T_\textrm{D}$ (K) & Ref. for $T_\textrm{c}$\\ 
    \hline \hline
    SrTiO$_3$  & electron (Nb, La, $V_{\textrm{O}}$) &  511-593 \cite{McCalla:2019} & \splitcell{This study \\ and \cite{Lin:2014,Collignon:2017,Thiemann:2018, Koonce:1967}} \\
    \hline
    C & hole (B) &  2240 \cite{Tohei:2006} & \cite{Bustarret:2015} \\ 
    \hline
    Si & hole (B) &  640 \cite{Bustarret:2015}&  \cite{Bustarret:2015} \\ 
    \hline
    SiC & hole (B, Al) &  1200 \cite{Bustarret:2015} &  \cite{Bustarret:2015} \\ \hline
    Ge & hole (Ga) &  360 \cite{Bustarret:2015} &  \cite{Bustarret:2015} \\ \hline
    SnTe & hole ($V_{\textrm{Sn}}$)& 140 \cite{Bustarret:2015} &  \cite{Bustarret:2015} \\ \hline
    GeTe & hole ($V_{\textrm{Ge}}$)& 170 \cite{Bustarret:2015} &  \cite{Bustarret:2015} \\ \hline
    PbTe & hole (Tl) &  140 \cite{Bustarret:2015}  & \cite{Bustarret:2015} \\ 
    \hline 
    Ba$_{\textrm{1-x}}$K$_\textrm{x}$BO$_{\textrm{3}}$ & hole & 210 \cite{Hundley:1989} & \cite{Pei:1990}  \\ 
    \hline
    A$_3$C$_{60}$ & electron & 2320 \cite{Gunnarsson:2004}  &  \cite{Gunnarsson:2004, Uemura:1991} \\ 
    \hline
    YNi$_2$B$_2$C & electron & 490 \cite{Michor:1995}  & \cite{Michor:1995}  \\ 
    \hline
    MgB$_2$ & electron & 750 \cite{Walti:2001}   & \cite{Nagamatsu:2001}\\ 
    \hline
    Bi & electron & 120 \cite{DeSorbo:1958}  & \cite{Prakash:2017} \\ 
    \hline
    Li$_\textrm{x}$ZrNCl & electron & 1044 \cite{Heid:2005}  & \cite{Taguchi:2006}  \\
    \hline
    RPdBi & electron &  295 \cite{Xu:2014}  & \cite{Nakajima:2015} \\ 
    \hline
    MA-TBG & electron/hole & 1861 \cite{Cocemasov:2015} & \cite{Cao:2018} \\ 
    \hline
    1ML-WTe$_2$ & electron &  130 \cite{Kimura:2019} & \cite{Fatemi:2018,Sajadi:2018} \\
    \hline
    
  \end{tabular}
  
\caption*{Table S1. Materials information for Fig. 5 in the main text. There, $E_\textrm{F}$ for all materials except SrTiO$_3$ is estimated from $E_\textrm{F}=\hbar^2/(2m^*)(3 \pi^2 n)^{2/3}$ for 3D materials and $E_\textrm{F}=\hbar^2/(2m^*)(2 \pi n_\textrm{2D})$ for 2D materials, where $n_\textrm{2D}$ is the 2D carrier density. }
\label{TableforTc}
\end{table}

\clearpage

%

